# Return on citation: a consistent metric to evaluate papers, journals and researchers

Tiancheng Li

BISITE research group, Faculty of Science, University of Salamanca, Salamanca 37008, Spain; and
School of Mechanical Engineering, Northwestern Polytechnical University, Xi'an 710072, China.
tiancheng.li1985@gmail.com, t.c.li@mail.nwpu.edu.cn
Webpage: https://sites.google.com/site/tianchengli85

**Abstract:**

Evaluating and comparing the academic performance of a journal, a researcher or a single paper has long remained a critical, necessary but also controversial issue. Most of existing metrics invalidate comparison across different fields of science or even between different types of papers in the same field. This paper proposes a new metric, called return on citation (ROC), which is simply a citation ratio but applies to evaluating the paper, the journal and the researcher in a consistent way, allowing comparison across different fields of science and between different types of papers and discouraging unnecessary and coercive/self-citation.

# I. Introduction

Assessing/evaluating or comparing the academic performance of a journal (or conferences), a researcher or a single paper (hereafter they are the same referred to as a "publication individual") has remained a critical, increasingly necessary and important issue with no standard solution where the pressure comes from finding a way to fairly measure the value of a researcher's published work. An efficient metric/measure should be simple, fair and positively meaningful. The principle of simplicity is important for the general use and easy understanding for the general public/researchers while on the contrary, to be fair it often needs to take into account various factors about the different nature of researches. It is challenging to achieve both simplicity and fairness but still, a good design would gain the maximum possibility.

A variety of metrics such as the known number of citations (primarily for the paper and the researcher), impact factor [1, 2] (for the journal) and H-index [3] (for the researcher) and their improvements [4] have been proposed. Almost all of them are based on the citation received (cited by what and how many times), which is the primary factor reflecting the quality of a publication but suffers from a common problem that they prevent comparing across different fields of science or even different types of papers. It is believed that publishing review/survey papers rather than standard research paper will be more helpful to increase the number of citations, impact factor and even h-index of a publication individual. What's worse, researchers may tend to work in hot disciplines/topics that can potentially produce more publications or attract more citations so that they can gain better academic scores. More seriously, some tend to (or be forced to) cite publications from particular organizations or persons including themselves (hereafter referred to coercive/self-citation [5, 6, 7]).

An efficient metric that is qualified to fairly evaluate the paper, the journal and the researcher either in a consistent way or across different disciplines and different types of papers, is still missing. For example, using the impact factor that is assigned to a journal to assess the quality of a paper published on that journal or the productivity of an individual researcher related to that journal is known to be highly questionable [1, 2]. Also, it is default to convincingly verify and prevent coercive/self-citation. So, is it possible to have a consistent metric that is able to solve all these problems?

This paper aims to propose a new metric that can directly apply to papers, journals and researchers, having clear and consistent meaning and allowing comparison across different disciplines and different types of publications. As the key idea, the number of citations made by the publication individual which on average reflects somewhat the differences between different disciplines and between different types of publications will be taken into account. The new metric is defined as a citation ratio, which is as simple as impact factor and is strongly complementary to existing metrics.

The paper is organized as follows. Section II presents the definition of the proposed new metric and its calculation. Section III discusses its advantage and disadvantage, where it is shown that the advantages is significant as compared with the slight disadvantage. The paper is concluded in Section IV.

## II. Return on citation (ROC): A novel metric and its calculation

The new metric as called "return on citation (ROC) [1]", is defined as a citation ratio as follows

$$\text{ROC} = \frac{\textit{The number of citations received}}{\textit{The number of citations made}} \quad (1)$$

As shown, the calculation of ROC of a publication individual comprises two parts, 1) the numerator, which is the number of times that the publication individual has been cited by other papers and 2) the denominator, which is the number of times that the publication individual has cited other papers.

The calculation of ROC is simple and straightforward[2]. However, it is possible, although it is very rare, that a publication make no citation, for which we suggest to take the denominator as 0.5 to enable the dividing calculation. Further on, one may exclude self-citations in the calculation of the numerator of (1) or give higher weight to citations from highly ranked journals. It is worth noting that when ROC applies to a researcher/journal, the calculation does not matter the type of publications that the researcher/journal has. Therefore, letters, commentaries, etc. that have not been counted in the calculation of impact factor[3] will also be included in the calculation of ROC.

As default, the number of citations received is calculated for all the past time, which gives the whole life performance so far. We note that the metric can also be specified for a particular period of the past when it applies to a journal or a researcher. The ROC of a researcher or a journal for a particular period gives the average performance of all the papers published in that period. For example, the ROC of a researcher or a journal in 2010 only count the corresponding papers published in 2010 by the researcher or journal, which is given by

$$\text{ROC}_{2010} = \frac{\textit{The total number of citations received by all the papers publised in } 2010}{\textit{The total number of citations made by all the papers publised in } 2010} \quad (2)$$

It is worth noting that, the denominator of (1) for a particular paper is fixed once the paper is published while it will grow with new publications appearing when the default ROC applies to a journal or a researcher. When the ROC of a journal or a researcher is specified for a particular period of the past, the denominator is also fixed. Therefore, we have the following conclusions:

**Remark 1**. The ROC of a particular paper will increase or not but will never decrease over time;

**Remark 2**. The ROC of a particular journal or a particular researcher can either increase or decrease over time;

**Remark 3**. The ROC of a particular journal or a particular researcher for a particular past period will increase or not but will never decrease over time.

---

[1] ROC is inspired by ROI (return on investment) used in business that is used to evaluate the efficiency of an investment or to compare the efficiency of a number of different investments.
[2] This involves the database used, which varies the number of citations much.
[3] Citations of commentaries, news & views articles, etc. contribute to the numerator of impact factor of a journal although these items are not counted in the denominator. This is unfair.

Regarding that the number of cites shall be approximately equal[4] to the number of times being "cited" among all the publications worldwide, we suggest to take 1 or lower[5] as a benchmark to assess whether a particular publication individual is "outstanding". The higher the ROC, the more outstanding the publication individual.

## III. Discussions: advantages, side-effects and beyond

For the detail, pros and cons of so many metrics proposed so far, the reader is referred to e.g. [1, 2, 3, 4, 5, 6] and the references therein. This paper is not intended to detail them, but we point out that the basis of almost all of them is the number of citations that the publication individual have received, whether in all time or in a specified period. But, no metric takes into account the number of citations that the publication individual makes, i.e. the number of times it cites others, which seems not directly related to the quality of the publication. However, it is helpful to distinguish between different disciplines and between different types of papers (reviews/surveys vs research papers) as working in a wide and hot discipline often have more related work to cite and a review paper also often cite more references than a normal research paper does. For this reason, the number of citations made by the publication individual is taken as a denominator factor in the calculation of ROC. This simple idea is indeed powerful as it provides a consistent metric for evaluation of papers, journals and researchers, allowing for comparison across different disciplines or between different types of papers and positively discouraging coercive/self-citation.

First, ROC can directly apply to a paper, a journal or a researcher and has always the same consistent and clear meaning. A high ROC means that the publication gains citation compare favourably to its citation, in terms of the number[6]. This performance measure does not necessary distinguish between a paper, a journal or a researcher nor between different disciplines, providing a consistent metric that is able to comprehensively evaluate all of them. Particularly, the ROC of a researcher, a journal or even a publisher is the average performance of all her/his/its publications as ROC is a citation ratio.

Secondly, as stated, one of the most known criticisms for impact factor and citations counts is that they invalidate comparison across disciplines and between reviews and research papers. ROC can simply alleviate this problem as the papers that seem to be able to attract more citations whether because they are in a hot disciplines or they are review often have cited more references accordingly. Then, the denominator of ROC will compensate for the difference caused by different disciplines or types of papers.

Thirdly, ROC will discourage unnecessary and coercive/self-citation as more citations to others will increase the denominator of ROC, reducing ROC. The authors may tend to only cite the very necessary work in order to get a high ROC. This is the good side but on the sad side, the misuse of any single metric can cause manipulation that the authors may become stingy to cite others even necessary. If the authors over-emphasize ROC, the average number of citations in the field will decrease. However, the adequacy of citations/references is a critical part for a paper, it is not a good idea to reduce the number of necessary references as it will impair/compromise the quality of the paper for attracting citations. Also, one of the peer reviewing criterions for many journals is based on the adequacy of citations. Therefore, we might not need to worry about much on the decreasing citations because of the use of ROC. However, we note that thorough and practical study about the impact of ROC is desired.

---

[4] Note that not all citations of publications go to another publication, but the citations may include webpages, patents, etc.
[5] Here, we are currently lacking of real data to determine a more rigorous benchmark level but we conjecture the levels across different disciplines will be close to each other.
[6] One can further consider about the quality of citations as well.

Furthermore, one can apply ROC to a publisher such as the association of "Nature" or "IEEE" by counting all the citations received and made by all the publications of the publisher in the calculation of a single ROC. This will give the academic performance of the publisher.

For a journal, a researcher or even a publisher, it might be interesting to know further about the statistics of the ROC performance of its individual papers such as 1) the largest individual ROC from all the papers published by a researcher/journal, 2) the number or proportion that the individual ROC of all the papers from a researcher/journal is larger than 1, as called High-ROC index/ratio ($No_{ROC>1}/R_{ROC>1}$), which can be calculated respectively as follows:

$$\text{ROC}_{max} = \max \text{ (ROCs of all the papers from a journal/researcher)} \quad (3)$$

$$\text{No}_{ROC>1} = \text{The total number of publications gaining a ROC larger than } 1 \quad (4)$$

$$\text{R}_{ROC>1} = \frac{\text{No}_{ROC>1}}{\text{The total number of publications}} \times 100\% \quad (5)$$

We reiterate that any single metric/measure has its limitations and it is fully not recommended to use a single metric to assess a publication individual, regarding to the diversity and multidimensionality of the nature of different research. Instead, a hybrid group of metrics in which one compensates another shall be used for fairer evaluation. The proposed ROC is simple, meaningful and strongly complementary with existing metrics, therefore we expect it to be a useful measure to be used jointly with existing metrics. Particularly, the idea of taking into account the number of citations made by the publication is critical and might be inspirable to design other metrics.

## IV. Conclusion

A new metric for evaluation the academic performance of a paper, journal or a researcher in a consistent manner has been proposed. The key idea is taking into account the total number of citations that the paper/journal/researcher makes in its/her/his publications. This provides several critical advantages over existing metrics such as allowing the comparison across disciplines and between reviews and research papers, discouraging unnecessary and coercive/self-citation. The proposed ROC is simple, meaningful and complementary with existing metrics, therefore it is useful to be used jointly with existing metrics.

The future work is using real data from papers, journals and researchers to test the validity and potential impacts of the proposed ROC metric.

## References

[1] http://en.wikipedia.org/wiki/Impact_factor, retrieved 24 December 2014.
[2] E. Garfield, The agony and the ecstasy — the history and the meaning of the journal impact factor, the 5th International Congress on Peer Review in Biomedical Publication, Chicago, USA, September 2005.
[3] http://en.wikipedia.org/wiki/H-index, retrieved 24 December 2014.
[4] M. Maabreh and I. M. Alsmadi, A survey of impact and citation indices: limitations and issues, International Journal of Advanced Science and Technology, vol. 40: 35-53, 2012.
[5] A. W. Wilhite and E. A. Fong, Coercive citation in academic publishing, Science, vol. 335 (6068): 542–543, 2012.
[6] B.D. Thombs, A.W. Levis, I. Razykov, A. Syamchandra, A.F. Leentjens, J.L. Levenson and M.A. Lumley, Potentially coercive self-citation by peer reviewers: A cross-sectional study, J Psychosom Res. vol. 78(1):1-6, 2015.
[7] T. Yu, G.Yu and M.-Y.Wang, Classification method for detecting coercive self-citation in journals, Journal of Informetrics, vol. 8(1):123–135, 2014.